\documentclass[9pt,twocolumn]{article}

\usepackage[paperwidth=8.5in, paperheight=11in,
            top=0.75in, bottom=1in,
            left=0.75in, right=0.75in,
            columnsep=0.33in]{geometry}
\usepackage{times}
\usepackage{graphicx}
\usepackage{booktabs}
\usepackage{xcolor}
\usepackage{hyperref}
\usepackage{amsmath,amssymb}
\usepackage{caption}
\usepackage[numbers,sort&compress]{natbib}
\usepackage{balance} 
\usepackage{adjustbox}

\usepackage{siunitx}
\DeclareSIUnit\flop{FLOP}
\DeclareSIUnit[per-mode=symbol]\floppersec{\flop\per\second}
\usepackage{minted}

\newcommand{\jupiter}[1]{\textcolor{black}{#1}}

\title{Computing the Full Earth System at 1~km Resolution}

\author{
Daniel Klocke$^{1}$ \and
Claudia Frauen$^{2}$ \and
Jan Frederik Engels$^{2}$ \and
Dmitry Alexeev$^{3}$ \and
Ren\'e Redler$^{1}$ \and
Reiner Schnur$^{1}$ \and
Helmuth Haak$^{1}$ \and
Luis Kornblueh$^{1}$ \and
Nils Br\"uggemann$^{1}$ \and
Fatemeh Chegini$^{4}$ \and
Manoel R\"ommer$^{5}$ \and
Lars Hoffmann$^{5}$ \and
Sabine Griessbach$^{5}$ \and
Mathis Bode$^{5}$ \and
Jonathan Coles$^{6}$ \and
Miguel Gila$^{6}$ \and
William Sawyer$^{6}$ \and
Alexandru Calotoiu$^{7}$ \and
Yakup Budanaz$^{7}$ \and
Pratyai Mazumder$^{7}$ \and
Marcin Copik$^{7}$ \and
Benjamin Weber$^{7}$ \and
Andreas Herten$^{5}$ \and
Hendryk Bockelmann$^{2}$ \and
Torsten Hoefler$^{7}$ \and
Cathy Hohenegger$^{1}$ \and
Bjorn Stevens$^{1}$
}

\date{
\small
$^{1}$Max Planck Institute for Meteorology, Hamburg, Germany\\
$^{2}$Deutsches Klimarechenzentrum, Hamburg, Germany\\
$^{3}$NVIDIA, Zurich, Switzerland\\
$^{4}$University of Hamburg, Hamburg, Germany\\
$^{5}$Forschungszentrum J\"ulich, J\"ulich, Germany\\
$^{6}$Swiss National Supercomputing Centre, Lugano, Switzerland\\
$^{7}$ETH Zurich, Zurich, Switzerland\\[0.35em]
\textit{Corresponding author:}
\texttt{daniel.klocke@mpimet.mpg.de}
}

\begin{document}

\makeatletter
\twocolumn[
  \begin{@twocolumnfalse}
    \maketitle
    \begin{abstract} \textbf{
    We present the first-ever global simulation of the full Earth system at \qty{1.25}{\kilo\meter} grid spacing, achieving highest time compression with an unseen number of degrees of freedom. 
    Our model captures the flow of energy, water, and carbon through key components of the Earth system: atmosphere, ocean, and land. 
    To achieve this landmark simulation, we harness the power of \num{8192} GPUs on Alps and \num{20480} GPUs on JUPITER, two of the world’s largest GH200 superchip installations. 
    We use both the Grace CPUs and Hopper GPUs by carefully balancing Earth’s components in a heterogeneous setup and optimizing acceleration techniques available in ICON’s codebase. 
    We show how separation of concerns can reduce the code complexity by half while increasing performance and portability. 
    Our achieved time compression of 145.7 simulated days per day enables long studies including full interactions in the Earth system and even outperforms earlier atmosphere-only simulations at a similar resolution.}
    \end{abstract}
    \vspace{1.0em}
  \end{@twocolumnfalse}
]
\makeatother
\maketitle

\section{Overview of the Problem} \label{sec:overview}

 Climate change is one of the greatest challenges facing societies and ecosystems. How human activities influence the climate depends on a complex interplay among diverse physical, biological, and chemical processes within the Earth system, whose non-linear interactions are coupled across a wide range of spatial and temporal scales. The varied processes, time, and spatial scales involved make their computation an extraordinary challenge. 
 
 At the heart of the challenge is the ambition to incorporate scales and processes necessary to physically couple the cycles of water, energy, and carbon -- the alpha and omega of climate science. Coupling the energy and water cycle requires resolving the gravest modes of convective instability, which most of us experience as weather. The qualitative change in a model able to resolve the dominant modes of moist convection explains why global scale simulations at \qty{1}{\kilo\meter} have long been the holy-grail for those interested in the physics of the climate and climate change~\cite{TomitaEtAl2005,ShuklaEtAl2010,Slingo2022}.
 
 Coupling the cycles of water and energy to that of carbon results in a manifestly more complex system. Flows of carbon among Earth's spheres depend on biogeochemical and dynamical processes in the ocean, and a cacophony of processes within the land biosphere \cite{friedlingsteinHowPositiveFeedback2003}. This then adds process complexity (biology, chemistry) to the scale complexity of the physical climate system. 

 Crowning these difficulties is the requirement that they be resolved without overly compromising system throughput. The temporal compression $\tau,$ expresses this throughput in units of simulated time versus actual time. Scientifically interesting studies are enabled by a $\tau$ of a few tens, to a few hundreds. With $\tau=30$ it becomes easy to study the diurnal cycle, and starts to become practical to study interannual variability, and hence Earth's main modes of forced variability. 
 
 With $\tau=300$ it becomes easy to study interannual variability, and multi-decadal simulations start to become practical. This enables studies of changes observed over the instrumental record; the exploration of scenarios of future changes to inform adaptation; or the investigation of irreversible changes that can have a bearing on mitigation efforts~\cite{ShawStevens2025}.

 In the past, researchers adopted one of two methods to circumvent an inability to explicitly incorporate scales and processes necessary to physically couple the cycles of water, energy, and carbon with sufficient temporal compression. One method was to restrict models to a particular component of the Earth system to capture scale interactions at short time-scales and / or over limited domains~\cite{SatohEtAl2019,ScharEtAl2020}. The second method was to forgo scale interactions, use statistical representations of fast and fine processes and include as many processes as possible of the full Earth system~\cite{BonyEtAl2013d,ShawStevens2025}. The former could take advantage of high-performance computing. The latter entailed a computing paradigm that favors simple and smaller computing architectures~\citep{ChenEtAl2021}. From a scientific perspective, by adopting the first approach, researchers decoupled the carbon from the water and energy cycles to understand water and energy. In the second approach, the water cycle was decoupled from the energy and carbon cycles to study energy and carbon. \emph{In this paper we are first to demonstrate how new heterogeneous computing architectures make it now possible to simulate the coupled water, energy, and carbon cycles all on a 1.25-km-scale global mesh and with a temporal compression that allows numerical experimentation on timescales of years to decades.}

 The size of the challenge of computing the coupling of water, energy, and carbon with a temporal compression sufficient to answer meaningful scientific questions can be appreciated by considering that a global, km-scale mesh with roughly 100 layers spanning the depth of the atmosphere, respectively the ocean, entails about \num{5.e10} spatial degrees of freedom -- or grid cells. When these are combined with prognostic quantities associated with these cells, e.g., the state of the atmosphere and ocean, plus tracers for H$_2$O, CO$_2$, and O$_3$ in the atmosphere, and 19 biogeochemical quantities in the ocean, properties describing the state of soils, the terrestrial biosphere (e.g., plant functional types and various carbon pools), and surface snow and ice, the result is roughly \num{1e12} physical-spatial degrees of freedom for our largest configuration (Table~\ref{tab_resolution}). Storing those degrees of freedom alone requires 8 TiB of main memory in double precision, more than the biggest announced trillion-parameter AI models. The particular challenge of simulating the Earth system at km-scale resolution then becomes one of taming the heterogeneity of a gargantuan system without compromising either the computational weak or strong scaling.

\begin{figure}[htp]
  \centering
  \includegraphics[width=0.9\linewidth]{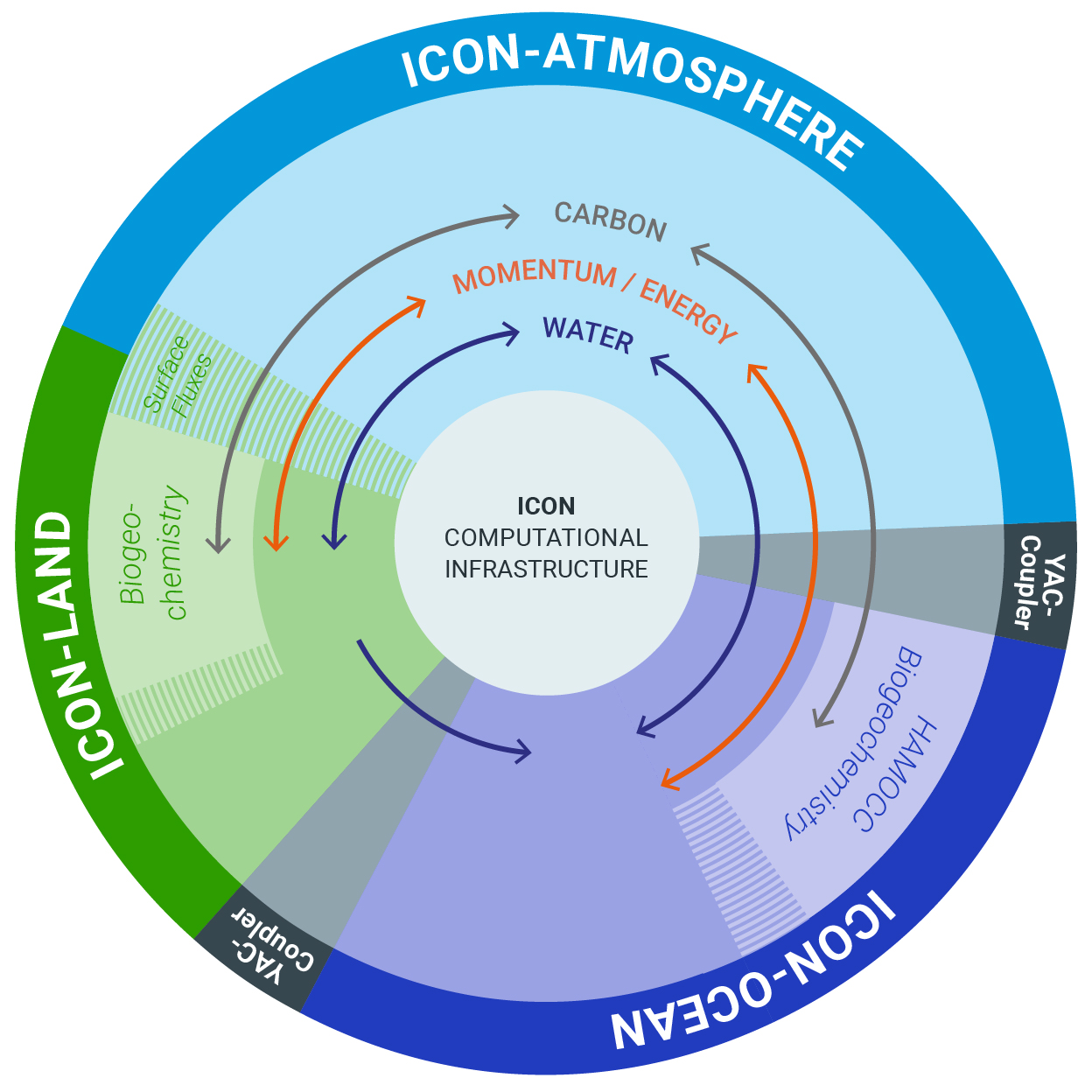}
  \caption{Schematic showing the main components of the Earth system in ICON and the exchange of energy, water, and carbon among them. Table~\ref{tab_resolution} presents the degrees of freedom assigned to each component.} 
  \label{fig:ICON-Modell}
\end{figure}

To meet this computational challenge requires harmonizing the representation of the system across a code base that incorporates an enormous range of processes operating at very different timescales with different spatial discretizations. For example, by virtue of its much greater mass, the ocean moves slowly, and can be marched forward with time steps five or ten fold longer than used by the atmosphere. Where as the computational effort in the ocean is dominated by a few large computational kernels, over the land (and to a lesser extent in the atmosphere) it is distributed over a large number of processing elements, posing challenges for multi-threading, especially on GPUs. Over the land and the atmosphere, the governing equations allow a local representation of the processes, while over the ocean a global solver is still required, thereby imposing different computational constraints. The relatively simple dynamics of each degree of freedom, and the strong coupling among degrees of freedom, means the final computations are not arithmetically intensive and hence memory bandwidth limited. The required harmonization of the code elements thus requires a meticulous management of an enormous flow of data through a complex computational architecture.

\section{Current State of the Art}

ICON \cite{icon2024} is a unified global numerical weather forecast and climate modeling framework. It uses a nonhydrostatic formulation of the atmosphere and performs its computations on a icosahedral-triangular C grid as described in~\citep{Giorgetta2018}. In the vertical it uses a terrain following hybrid sigma height grid for the atmosphere~\citep{Leuenberger2010}. It includes components for atmosphere, land, the terrestrial biosphere (vegetation), ocean, and sea ice, and ocean biogeochemistry (see \autoref{fig:ICON-Modell}). It is mostly written in Fortran and uses a hybrid MPI/OpenMP approach for parallelization. Additionally, the atmosphere and land components have been ported to GPUs using OpenACC~\citep{Giorgetta2022}.

The configuration of ICON benchmarked for this study has no counterpart. It unifies two separate approaches (\S\ref{sec:overview}) that, until now, have defined the state of the art. These are: (i) simulations that are performed to resolve fine and fast scales globally for single components of the Earth system on state-of-the-art computing facilities; and (ii) complex Earth system models that are run by approximating the behavior of fast and fine scales with statistical models, configurations which then do not benefit from high-performance computing.

\begin{table*}[htp]
\caption{\sisetup{detect-all=true}Overview of current state of the art km-scale climate simulations, their horizontal grid spacing, included Earth system components (A - Atmosphere, L - Land, V - Vegetation, O - Ocean, B - Biogeochemistry, C - Carbon), the system they ran on, the temporal compression $\tau$ (e.g., Simulated Days Per Day) and a rescaled value  $\tau^{*}_{1.25}$ corresponding to the expected performance on the same resource but with $\Delta x = \qty{1.25}{\kilo\meter}$.}
\label{tab_tobeat}
\begin{center} 
 \begin{tabular}{lrrlrcrr}
 \toprule
 Model & $\Delta x$ / km &  Components & Resource & $\tau$ & $\tau^*$ \\
\midrule
 SCREAM \citep{Taylor2023}          & 3.25/4.875   & \texttt{A L - - - -} & \qty{\approx87}{\percent} Frontier GPU    & 458 & 26      \\ 
 ICON     & 1.25         & \texttt{A L - O - -} & \qty{\approx95}{\percent} Lumi GPU        & 69 & 69         \\ 
 NICAM (H. Yashiro pers.)           & 3.5          & \texttt{A L - - - -} & \qty{\approx26}{\percent} Fugaku CPU     & $\approx 365$ & 17   \\ 

\midrule
 \textbf{this work} & \textbf{1.25}  &  \textbf{\texttt{A L V O B C}} & \qty{\approx85}{\percent} JUPITER GPU & \textbf{145.7} & \textbf{145.7} \\
  
\bottomrule
\end{tabular}
\end{center}
\label{tab:competion}
\end{table*}

Two recent benchmarks define the high-performance frontier for Earth system simulations. The Simple Cloud-Resolving E3SM Atmosphere Model (SCREAM) was implemented from scratch using C++ and Kokkos. Its highest resolution was configured with $\Delta x =\qty{3.25}{\kilo\meter}$ and 128 vertical levels (with a somewhat coarser, $\Delta x = \qty{4.875}{\kilo\meter}$ physics grid). SCREAM achieved a $\tau=458$ by scaling across \num{8192} nodes -- \num{32768} Instinct MI250X GPUs -- on Frontier~\cite{Taylor2023}, an accomplishment recognized by the inaugural Gordon Bell Prize for Climate Modeling. At a $\Delta x=\qty{3.5}{\kilo\meter},$ similar to the SCREAM benchmark, a Fortran implementation of the atmosphere model NICAM has reported $\tau\approx365,$ for a simulation on \num{40960} nodes or \qty{26}{\percent} of Fugaku (H. Yashiro personal communication). The different temporal compressions of the applications can be roughly related to each other by
calculating a reference $\tau,$ which we call $\tau^*.$ It rescales the $\tau$ from previous applications to estimate its value were the calculations performed with $\Delta x = \qty{1.25}{\kilo\meter}$ on the same resource, i.e., $\tau^* = (1.25/\Delta x)^3 \tau_{\Delta x}.$  This yields $ \tau^*$ of 17 and 26 for both NICAM and SCREAM, respectively (Table \ref{tab_tobeat}). At even finer scales, a $\Delta x =\qty{0.22}{\kilo\meter}$ version of NICAM has has been run across half of Fugaku~\cite{MatsugishiEtAl2024}, but its performance characteristics are not available.

Other components than the atmosphere of the Earth system have greater throughput and more varied coding paradigms.  For instance, global ocean models have been developed using a variety of programming paradigms, including Python with a JAX backend~\cite{Hafner2021} and Julia~\cite{wagner2025EtAl}, with $ \tau^* \approx 500$ being reported~\cite{wagner2025EtAl}. Over the land surface, continental scale simulations of ground water flow at $\Delta x = \qty{1}{\kilo\meter}$ have been performed and while performance measures are not given, the ability to iterate them to steady state suggests even larger values of compression~\citep{Maxwell2015}. For both the ocean and for the atmosphere, high-resolution simulations over regional domains are common~\cite{GulaEtAl2016,RasmussenEtAl2023}, but because $\tau$ is determined by the finest scales, regional models do not introduce new computing challenges. With the advent of machines like Alps, Fugaku, Frontier, or the emerging JUPITER, $\tau$ has ceased to be restricted by domain size for global domains with $\Delta x > \qty{500}{\meter}.$ Hence looking across component models it appears that the fundamental roadblock is the scaling of the global atmosphere component, for which the state of the art is $ \tau^*\approx 20,$ and the important challenge of tying all the components together in a heterogeneous computing environment.

\begin{figure*}[!ht]
  \centering
  \includegraphics[width=0.95\linewidth]{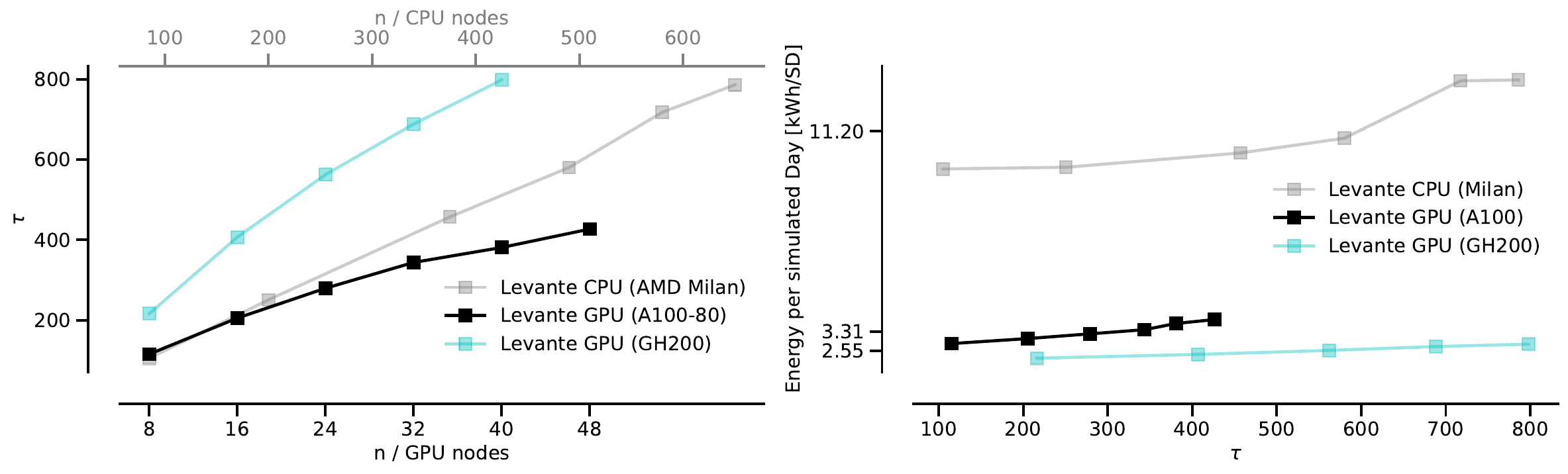}
  \caption{Strong scaling on CPUs and GPUs of the coupled 10~km ICON (without biogeochemistry submodel) on Levante (left). Nearly four-fold improved energy efficiency on Levante GPUs vs CPUs (right).}
  \label{fig:cpu-gpu}
\end{figure*}

Fully-coupled Earth system models are typically run at much coarser resolution and with less computational power. For the carbon cycle the consensus has been that even simulations at $\Delta x = \qty{10}{\kilo\meter}$ are out of reach~\cite{Roberts2025}. One reason for this has been the interest in using them to explore centennial scale variability, which only becomes practical with $\tau\approx\num{3000}$. 
To compare GPU and CPU performance and to also measure the energy consumption for running on GPUs vs running on CPUs we did additional simulations with a coupled atmosphere-ocean 10~km configuration on the Levante supercomputer (Figure~\ref{fig:cpu-gpu}). For ICON we achieve about a factor of 2 less throughput on the A100 nodes of Levante compared to the GH200 nodes.
On GPUs, strong scaling begins to decline at $\tau\approx798$ on 40 GH200 nodes. Reducing resolution by a factor of ten could, with perfect weak scaling, increase $\tau$ by a factor of ten but would not provide enough work for a GPU node. This puts a practical limit of about $\tau=\num{3192}$ and $\Delta x = \qty{40}{\kilo\meter},$ which could effectively utilize 2.5 GH200 nodes. 
In comparison, on the CPU nodes ($2\times$AMD 7763) strong scaling starts leveling off at a larger $\tau$ albeit at a much higher node count (see also~Mauritsen et al.~\cite{MauritsenEtAl2022}), and less dramatically as increased cache efficiency partially offsets the lack of computation. This makes it possible to dial the resolution back further and achieve a larger $\tau,$ without making the computation too small. This however comes at a considerable cost in power, with time to solution demanding 4.4 times as much power on CPUs (Figure~\ref{fig:cpu-gpu}, right). Which explains why, until now, Earth system models tend to be run at coarser resolutions on older computational architectures with smaller node counts. An additional factor is that a model with $\Delta x=\qty{40}{\kilo\meter}$ must parameterize (statistically represent) many processes that a model running at finer resolution can represent explicitly. This makes the model much more complex and its individual components difficult to separate and unit test. The model becomes more challenging to adapt to different computing paradigms even if doing so were computationally advantageous.

The first atmosphere-ocean-coupled km-scale models running at km-scale were developed by the group at the Max Planck Institute for Meteorology, working with the German Climate Computing Center in Hamburg~\cite{MauritsenEtAl2022,Hohenegger2023}. They initially reported $\tau=20$ for $\Delta x=\qty{2.5}{\kilo\meter}$ on \num{600} Levante CPU nodes. First simulations were performed for a few simulated months, later multi-decadal simulations have been conducted with a $\Delta x$ of \qty{5}{\kilo\meter} and \qty{10}{\kilo\meter}~\cite{egusphere-2025-509}. In parallel, a version of the ECMWF's IFS atmosphere model has been coupled to two different ocean models at horizontal grid-spacing of up to $\qty{4.4}{\kilo\meter}$~\cite{Rackow2022} for which they achieved a $\tau \approx 100$ on 100 nodes of Levante.

Recently a \qty{5}{\kilo\meter} configuration of ICON was used to simulate the annual cycle with ocean biogeochemistry included~\cite{Nielsen2025}. While still being considerable simpler, this version comes closest to the configuration we describe here.  Because it used the same code base as used here and its performance characteristics were not explored, it is not included in Table~\ref{tab_tobeat}.

\section{Innovations Realized}

Our main innovations that enable this work are arranged along two main axes: (1) exploiting functional parallelism by efficiently mapping components to specialized heterogeneous systems and (2) simplifying the implementation and optimization of an important component by separating its implementation in Fortran from the optimization details of the target architecture.

\subsection{Exploiting the Earth System's Complexity to Flexibly Utilize System Parallelism}
\begin{figure}[h!]
  \centering
  \includegraphics[width=0.75\linewidth]{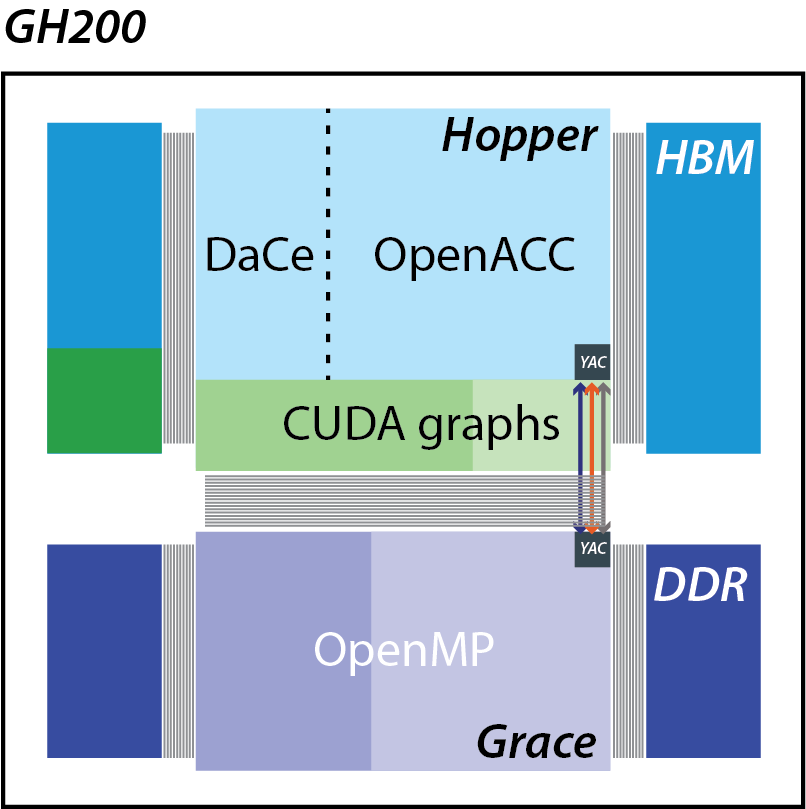}
  \caption{Mapping of the Earth system components (from Figure~\ref{fig:ICON-Modell}) to the GH200 superchip with indicated innovations. On the platforms tested, four superchips are on each node. For our benchmarking experiments, we utilized 10 MPI processes with each 6 OpenMP threads for the ocean on the CPUs and 1 MPI process for the GPU of each GH200 superchip.
  }
  \label{fig:GH200}
\end{figure}

The additional challenge of simulating the Earth system (as opposed to just the coupled atmosphere and ocean) is that a great number of additional components need to be modeled and their computation needs to be harmonized (see Table~\ref{tab:components}). What makes the problem challenging is that these components have different performance characteristics and none of them dominates the runtime. This leads to the well-known flat performance profile of Earth system models which makes optimization a painstaking task. To address this, we adopted different strategies, as illustrated schematically in Figure~\ref{fig:GH200}.

For the \textbf{atmosphere component}, two approaches were tested. Originally, the entire atmosphere was GPU-accelerated with OpenACC \citep{Giorgetta2022}. For this project we also replaced the OpenACC implementation of one of the largest kernels with a DaCe~\cite{dace} implementation, as described further below. Both implementations allow us to keep the data resident on the GPU throughout the whole simulation and avoid memory transfers between the GPU and the CPU, which can be prohibitively expensive on some architectures. The boundary exchange is done via MPI point-to-point communications, which is implemented in a way that it can employ direct GPU-GPU transfers with GPUDirect RDMA. \emph{The atmosphere component is highly optimized for parallel GPU execution as we will elaborate later.}

For the \textbf{land component} (JSBach), the introduction of an interactive biosphere model introduced a very large number of additional small GPU kernels. The small kernel problem was exacerbated by the small number of soil levels (5), which results in little computational intensity. Launching many small kernels on the GPU incurs a significant latency overhead and becomes very expensive. An innovative way to reduce this cost is to employ CUDA Graphs for the OpenACC kernels. CUDA Graphs record the kernel call flow of a component (like JSBach) and replay it exactly the same way in subsequent calls to this component. Recording causes a slightly increased latency for the first invocation, however, all the subsequent replays are virtually latency-free. Moreover, due to the JSBach model implementation operating on multiple independent plant functional types, many of the executed kernels may run simultaneously using CUDA graphs. \emph{Overall, we observe a speedup for the land and vegetation parts of the model on the order of 8-10$\times$ depending on the grid-spacing.}

While the \textbf{ocean component} has been ported with OpenACC to run on GPUs, it poses a special challenge for efficient scaling, especially on GPUs. This arises because filtering of fast wind-driven surface waves introduces a tightly-coupled 2d-equation-system distributed over all ranks to solve for its barotropic mode. The computational characteristic of this solver is dominated by global communication, while the computations in between communication are very small. In contrast to the land model, which directly exchanges fluxes with the atmospheric component on the atmospheric timestep, and therefore needs to run on GPUs (for most architectures), the slow evolution of the ocean allows it to be much more loosely coupled to the atmosphere and thus run concurrently to the atmosphere and land components. Only energy, water and carbon are exchanged between the atmosphere and the ocean at a coupling timestep every 10 simulated minutes through the coupler YAC \citep{Hanke2016}. These properties -- a global solver, loose coupling, and a much larger $\tau$ -- make the ocean very well suited to exploit CPU resources. Hence we take advantage of this flexibility to utilize the hybrid computational power of both CPUs and GPUs on the same nodes when available. Specifically, for the GH200 superchip we run the ocean on the Grace CPUs which are a powerful resource that would otherwise be underutilized. \emph{Given that most other simulations rely mostly on GPUs, by mapping the Earth system workload smartly, we essentially run the ocean component for free.}

\begin{table}[htb]
\centering
\small
\setlength{\tabcolsep}{4pt} 
\caption{Earth system model global grid configurations for ICON, showing the number of grid cells, vertical level and prognostic variables for each component model used in the scaling runs. Nominal resolution for ICON is given as the square root of the mean cell area. Velocity components normal to the triangle edges are counted as 1.5 prognostic variables. Land has four physical state variables on five levels, 21 additional carbon pools, plus the leaf area index as predicted variables, associated with up to 11 plant functional types (reported in the levels column for vegetation). Degrees of freedom are calculated as the product of the spatial degrees of freedom times their number of prognostic variables.}
\label{tab_resolution}

\begin{adjustbox}{width=\columnwidth}
\begin{tabular}{llrrrrr}
\toprule
 &  & $\Delta x$ / km & cells $\times10^8$ & levels & vars & $\Delta t$ / s\\
\midrule
\multicolumn{7}{l}{\textbf{10 km:} \num{1.2e10} deg. of freedom}\\
& Atmosphere               & 10   & 0.05  & 90        & 12.5  & 75  \\
& Land                     & ``   & 0.015 & 5         & 4     & ``  \\
& Vegetation               & ``   & ``    & $\leq$ 11 & 22    & ``  \\
& Ocean \& sea-ice         & ``   & 0.037 & 72        & 5     & 600 \\
& Biogeochemistry in ocean & ``   & ''    & ''        & 19    & ``  \\
\midrule
\multicolumn{7}{l}{\textbf{1.25 km:} \num{7.9e11} deg. of freedom}\\
& Atmosphere               & 1.25 & 3.36  & 90        & 12.5  & 10  \\
& Land                     & ``   & 0.98  & 5         & 4     & ``  \\
& Vegetation               & ``   & ``    & $\leq$ 11 & 22    & ``  \\
& Ocean \& sea-ice         & ``   & 2.38  & 72        & 5     & 60  \\
& Biogeochemistry in ocean & ``   & ''    & ''        & 19    & ``  \\
\bottomrule
\end{tabular}
\end{adjustbox}
\label{tab:components}
\end{table}

The \textbf{biogeochemistry component} of the ocean, HAMOCC, involves a large number of tracers (prognostic variables in Table~\ref{tab_resolution}) that interact with one another and are transported through the ocean. HAMOCC does not have a global solver, but shares the large $\tau$ of the ocean (compared to the atmosphere) due to the slow transport, and the loose coupling with the fast atmosphere. This makes HAMOCC better adapted for GPUs than the ocean dynamical core. To exploit more parallelism and gain flexibility in its implementation, a concurrent version of HAMOCC has been developed~\cite{Linardakis2022}, which allows it to be run on GPUs through OpenACC instrumentation of the Fortran code. A downside of this approach is that large three-dimensional fields need to be exchanged between the ocean dynamical core and HAMOCC at every ocean timestep. Therefore exploiting concurrent GPU parallelism in HAMOCC is not beneficial in all cases. For a powerful enough CPU like on the GH200 superchip it is also possible to include the biogeochemistry together with the ocean on the CPU, as it shares two of the features (loose coupling and larger $\tau$) and thus can more efficiently use a resource that would otherwise be idle. \emph{In our setup, we run HAMOCC together with the ocean component on the CPU and also essentially get it for free, concurrent to the GPU execution.}

\subsubsection{Final Mapping}
Summing up, our setup is as follows: For the atmosphere and land models we use four MPI ranks per node, each assigned to a GPU and pinned to the CPU cores of the same superchip. These ranks launch kernels on the GPU and can use a number of helper threads. In addition, for the ocean, sea ice, and biogeochemistry, we use 40 MPI ranks with six OpenMP threads each per node, ten ranks each pinned to its respective NUMA domain. Some of these ranks could also be utilized for asynchronous output. In total, we use 244 of the 288 available cores per node leaving the remaining cores available to support GPU and operating system services and reduce noise~\cite{noise-sim}.

To maximize throughput with our approach of utilizing all components of the system we require a delicate load balancing. Since the GPUs are significantly more powerful than the CPUs, we aim to keep the GPUs running at full capacity at all times. This requires the ocean component to consistently run faster than the atmosphere, ensuring that the ocean component rather than the atmosphere waits at the synchronization points where the coupler exchanges fluxes between the components. However, assigning too many CPU resources to the ocean to make it arbitrarily fast can actually slow down the atmosphere. An important advantage of the GH200 superchips is their ability to flexibly partition power consumption between CPUs and GPUs, which enables further performance optimization of heterogeneous configurations for specific node counts.

This is due to a key characteristic of the systems targeted by this project: CPU and GPU components share a common power and thermal budget -- \qty{660}{\watt} per superchip on Alps and \qty{680}{\watt} per superchip on \jupiter{JUPITER} (see Table \ref{tab:systems}) -- which is substantially lower than the maximum combined power capacity of both components (\qty{1000}{\watt}, according to NVIDIA). As a result, power distribution must also be carefully balanced, adding even more complexity to the setup. Fortunately, climate models -- and especially the ICON atmosphere component -- are predominantly memory-bandwidth bound, meaning the GPU does not require its full theoretical power budget, which is largely used by the compute facilities of the device.

\subsection{\hspace{-0.5em}Separation of Concerns: From Fortran to GPU}

To achieve performance and scalability on today’s heterogeneous systems, ICON’s code base has evolved to support a wide range of configurations: from pragmas enabling OpenMP and OpenACC parallelization, to Intel, Cray, GCC, and NEC vector annotations to macros that duplicate code to reorder loops. For example the dynamical core, the most compute-intensive part of the most expensive atmosphere component, contains \num{2728} non-empty Fortran source lines of code, of which less than \qty{50}{\percent} are actually describing the computation. The remaining lines of code are optimizations for specific architectures implemented using OpenACC (\qty{20}{\percent}) and other directives (\qty{12}{\percent}) as well as duplicated code reordering loops (\qty{6}{\percent}). The following figure shows a small code excerpt from ICON's dynamical core illustrating three different families of pragmas and how macros and duplicate loop headers are used to specialize for different architectures.

\vspace{0.5em}
\hrule
\vspace{0.5em}
\hspace{-1.2em}\includegraphics[width=\linewidth]{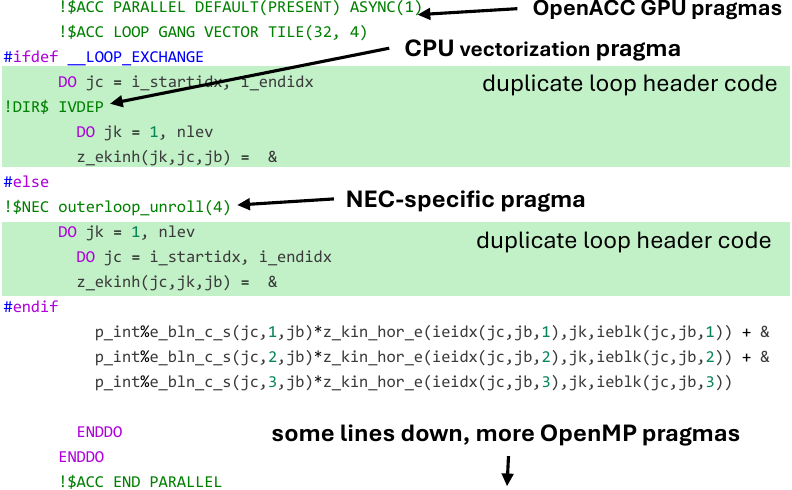}
\vspace{0.5em}
\hrule
\vspace{0.5em}

The resulting duplication makes not only the initial development costly but makes extending or maintaining the code excessively complicated. For example, changing a loop would require updating all copies consistently. Updating any source code would need to also update the describing pragmas accordingly. Yet, the systems described by those annotations may not even be available to the programmer at the time of the change.

We recognize that the Fortran language itself specifies the computations in a way that allows us to extract their dataflow dependencies. We extend DaCe~\cite{dace}, a parallel datacentric compilation framework, with a Fortran language parser specialized to the dynamical core code. We then remove all pragmas and macros from the code to create the cleanest form with only about \num{1400} lines of code (less than \qty{50}{\percent}). 

We use our new parser to read this sequential Fortran code into DaCe's internal "Stateful Dataflow Graph" (SDFG) representation~\cite{dace}, maintaining all of the code's semantics while representing data movement explicitly. This representation naturally enables us to analyze independent pieces of code and loops and extract the maximum parallelism, which we can then map again to different target architectures such as GPUs or CPUs. 

DaCe has been demonstrated to optimize the established weather and climate code FV3 after it has been rewritten in a Python dialect~\cite{fv3}. The innovation in this work is that \emph{we use the unmodified sequential Fortran code of ICON to achieve a similar result. This demonstrates a clear separation of concerns between the application scientist writing her Fortran code and the performance engineer using transformations and code-generation to accelerate it on various architectures. }

To achieve this separation of concerns, DaCe allows the performance engineers to write "performance metaprograms" that transform a piece of a SDFG into a new representation targeted at specific devices. We designed and wrote a series of such metaprograms to transform the dynamical core code into a fast representation for GH200 GPUs as well as CPUs. We note that all the optimizations are without changes to the original Fortran code and thus invisible to the climate scientist. If the Fortran code is changed but maintains a similar computational structure, then the transformations continue to apply as they are matched to dataflow structures on the SDFG, see Ben-Nun et al.~\cite{dace} for details.

We illustrate one code-specific optimization out of many that we applied: The ICON dynamical core performs its computations on an icosahedral grid composed of triangles and its dual grid composed of mostly hexagons and 12 pentagons. This structure requires a significant amount of index lookups in arrays to determine grid indices. Some of these indices can be reused by carefully reordering computations~\cite{indexreuse}. By leveraging the freedom offered by our maximally parallel representation, we can reduce the number of integer index lookups required per grid point by an average factor of 8$\times$.

After our transformations, the whole dynamical core code transformed via DaCe consistently outperforms the manually tuned OpenACC implementation - starting from the sequential Fortran implementation and requiring no pragmas or manual data layout transformations. The following figure shows an overview of the achieved performance for the 10 km setup comparing the version with manual pragmas (blue) to the DaCe implementation (orange) for the GPU execution on GH200.

\vspace{0.5em}
\hrule
\vspace{0.5em}
\hspace{-1.2em}\includegraphics[width=\linewidth,clip,trim={0.05cm 7.95cm 0cm 0cm}]{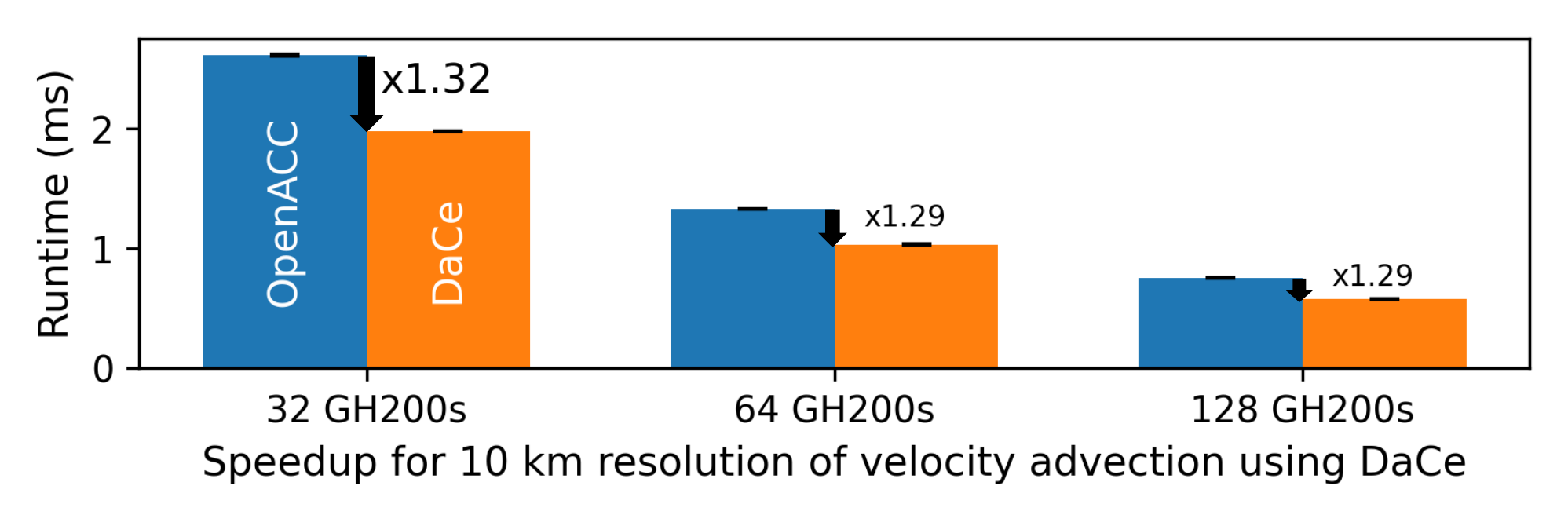}
\vspace{0.5em}
\hrule
\vspace{0.5em}

As the dynamical core is memory-bound, we evaluate the sustained bandwidth during execution for both the OpenACC and DaCe implementations. Assuming that \qty{100}{\percent} busy DRAM would yield a bandwidth of \qty{4}{\tebi\byte\per\second} on GH200 GPUs, we convert the DRAM utilization percentages reported by Nsight for these kernels to average bandwidth in different experiment configurations as is shown in the following figure. 
\vspace{0.5em}
\hrule
\vspace{0.5em}
{\centering
\includegraphics[width=\linewidth,clip,trim={0.05cm 6.80cm 0cm 0cm}]{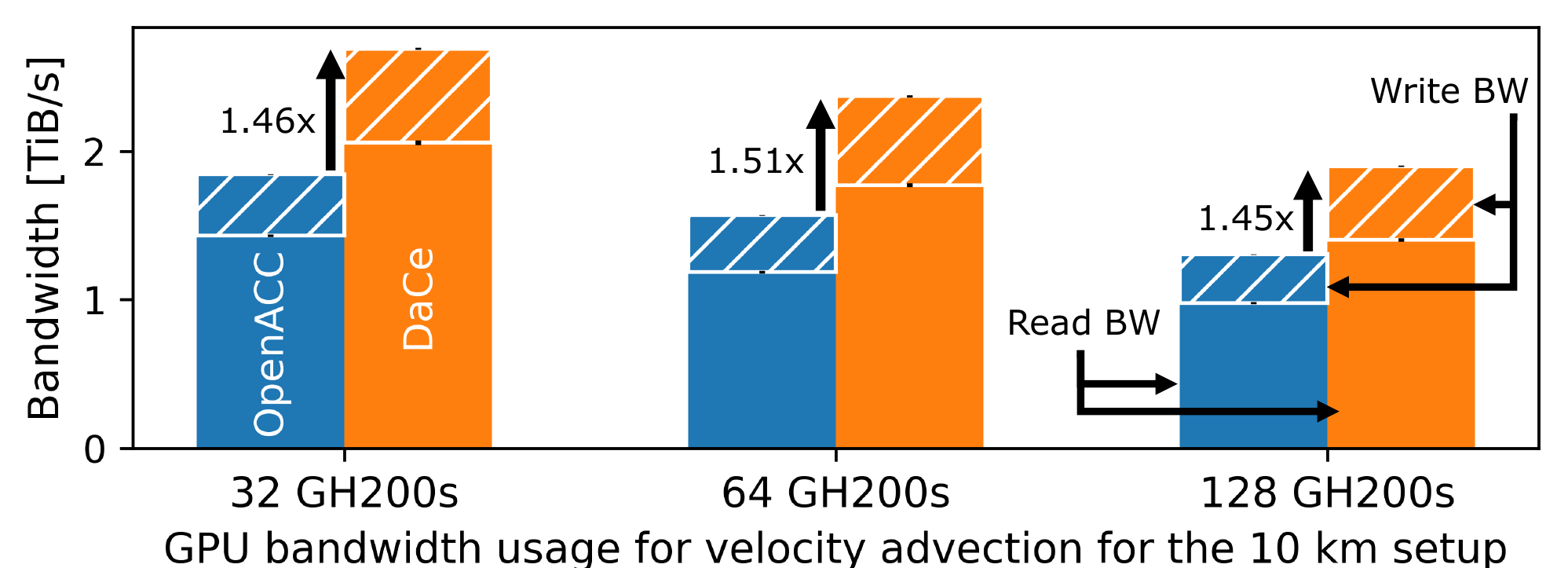}}

\hrule
\vspace{0.5em}
As the 128 GH200 configuration reflects the same work per chip as our hero run (see Figure~\ref{fig:StronAndWeak-Scaling}), it would achieve more than 15 PiB/s sustained memory bandwidth, which is about \qty{50}{\percent} peak.

\section{How Performance Was Measured}

\subsection{Earth System Model Configuration}
Computational benchmarks were performed using the coupled Earth system model ICON including the atmosphere, land (with dynamic vegetation), ocean, sea-ice, ocean biogeochemistry, hydrological discharge from land to ocean, and the carbon exchange between the components. The model was initialized with the Earth system state on the 1st of January 2020. For the ocean and its biogeochemical state a separate spin-up simulation was performed using prescribed surface conditions from reanalysis of the atmosphere. The initial atmospheric state was interpolated from reanalysis onto the computational grid and carbon-pools were initialized from separately spun-up states. For our simulations, we consider a development configuration with \qty{10}{\kilo\meter} horizontal resolution and a configuration to run at scale with \qty{1.25}{\kilo\meter} horizontal resolution. Both configurations only differ in their horizontal grid and the timestep length.

\subsection{System Details}\label{sec:systems}
The multi-component setup of ICON presented in this work is enabled by NVIDIA's Grace Hopper GH200 superchip in the Alps and JUPITER systems. The superchip combines in one package an ARM CPU, \emph{Grace}, with 72 cores and \qty{120}{\giga\byte} LPDDR5 memory, and a \emph{Hopper} GPU with \qty{96}{\giga\byte} HBM3 memory. Both parts are connected by the \emph{NVLink-C2C} bus with \qty{900}{\giga\byte\per\second} bandwidth, allowing high-bandwidth, cache-coherent access to the two memories~\cite{gh200}. The CPU and the GPU share one combined Thermal Design Power (TDP), in which power is dynamically distributed first to the CPU and the remainder to the GPU. The TDP value is system-dependent.

\begin{table}[htb]
\centering
\small
\setlength{\tabcolsep}{4pt} 
\caption{High-performance computing systems used.}
\label{tab:systems}

\begin{adjustbox}{width=\columnwidth}
\begin{tabular}{lrr}
\toprule
 & \textbf{JUPITER} & \textbf{Alps} \\
\midrule
\# Nodes & \num{5884} & \num{2688} \\
\# Superchips per Node & 4 & 4 \\
\# Superchips & \num{23536} & \num{10752} \\
Superchip TDP & \qty{680}{\watt} & \qty{660}{\watt} \\
Interconnect & InfiniBand NDR200 & Slingshot-11 \\
Inj. Bandwidth / Node & 
\qtyproduct[product-units=single, per-mode=symbol]{4 x 200}{\giga\bit\per\second} &
\qtyproduct[product-units=single, per-mode=symbol]{4 x 200}{\giga\bit\per\second} \\
\bottomrule
\end{tabular}
\end{adjustbox}
\end{table}

\subsubsection*{\jupiter{JEDI/JUPITER}}

JUPITER, which is currently being installed at the J\"ulich Supercomputing Centre in Germany, is expected to be the first European (HPL) exascale computer. The system currently consists of \num{5884} nodes (\num{23536} GH200 superchips), connected with NDR200 InfiniBand.
ICON is one of the driving applications for the design of JUPITER~\cite{10793128}.
Through early access we used \jupiter{JEDI}, a single rack (48 node) development platform, to prepare larger runs on Alps \jupiter{and then JUPITER}.

The code and setup we demonstrate here will be used in production to simulate 13 months at \qty{1.25}{\kilo\meter}, an unprecedented scale for the simulation of the fully coupled Earth systemx.

\subsubsection*{Alps}

Alps is a supercomputer operated by the Swiss National Supercomputing Centre CSCS that deploys \num{2688} nodes with the total of \num{10752} GH200 superchips to reach \qty{435}{\peta\floppersec} HPL performance on the TOP500 list (place 7; November 2024). The interconnect is Slingshot-11. 
An overview of both systems is presented in \autoref{tab:systems}.

\subsubsection*{Software Infrastructure}

On JUPITER, CUDA 12.6, NVIDIA HPC SDK 25.5, and OpenMPI 5.0.5 with support for CUDA-aware MPI were used.
Alps supports a very similar environment with Cray-MPICH instead of OpenMPI. The final runs on Alps used CUDA 12.8, NVIDIA HPC SDK 25.3 and Cray-MPICH 8.1.32. Because ICON runs so close to the machine limits, it is sensitive to the balance of many low-level system parameters at different node counts.  Moving from JEDI, where the initial development happened, to Alps, involved the close cooperation of the operators to harmonize these settings and maintain good performance across the entire system.

\subsection{Performance Metrics}

The most relevant performance metric for climate simulations is the temporal compression $\tau,$ which describes the model throughput in units of simulated time versus actual time. We derive $\tau$ from the time it takes to simulate 3 hours in the \qty{1.25}{\kilo\meter} configuration (24 hours in the 10~km configuration) excluding the initialization time. Initialization takes a measurable amount of time in our short benchmark runs. However, in a production setup when the model is run for hours its contribution to the overall run time is relatively small, which is why we exclude it here.
The simulation time is measured independently for the atmosphere/land and ocean/sea-ice/biogeochemistry components.
Included in timings is the coupling time, i.e., the amount of time atmosphere/land have to wait for ocean/sea-ice/biogeochemistry and vice versa. Ideally, the coupling time should be close to zero for the computational expensive atmosphere/land components running on the GPUs, meaning that they do not have to wait for the ocean/sea-ice/biogeochemistry components to finish their calculations. Timings are measured based on the CPUs time stamp counter. Measurements are taken across all ranks and the maximum value is presented.

\subsection{I/O Performance Metrics}

ICON has a number of different I/O schemes for various use-cases. Most notable are asynchronous schemes for model output, output via the coupling interface, writing and reading of restarts, and distributed reading of initial and boundary conditions.

Checkpoint/restart means that the model dumps the complete state from memory to disk. Since the state is huge, efficient and fast I/O is necessary. Checkpoint/restart I/O in ICON has a number of modes, from rank-0-I/O to asynchronous multi-file-I/O. We employ the synchronous multi-file variant, where a configurable subset of ranks collects the variables and writes them to one file each. Reading, in turn, can be done with a different subset of ranks, where each rank reads parts of the files and distributes the data to the corresponding ranks. 

With the asynchronous output scheme, additional MPI tasks are assigned to output servers. Output fields are communicated to these servers via MPI one-sided remote memory access in configurable intervals, additional operations (averaging, accumulating, interpolation to different output grid in vertical and/or horizontal) can be applied. Disk I/O takes place concurrently to the model integration. The overhead in computational resources depends on the number of fields and desired storage frequency. The required bandwidth to the storage is low and impact on neighbor exchange is small, so that I/O does not appreciably impact $\tau.$ Moreover, this is something that is aided by increased temporal sparsity output as $\Delta x,$ and hence the timestep, becomes small, while the output frequency remains constant.

\section{Performance Results}

\begin{figure*}[htb]
  \centering
  \includegraphics[width=0.95\linewidth]{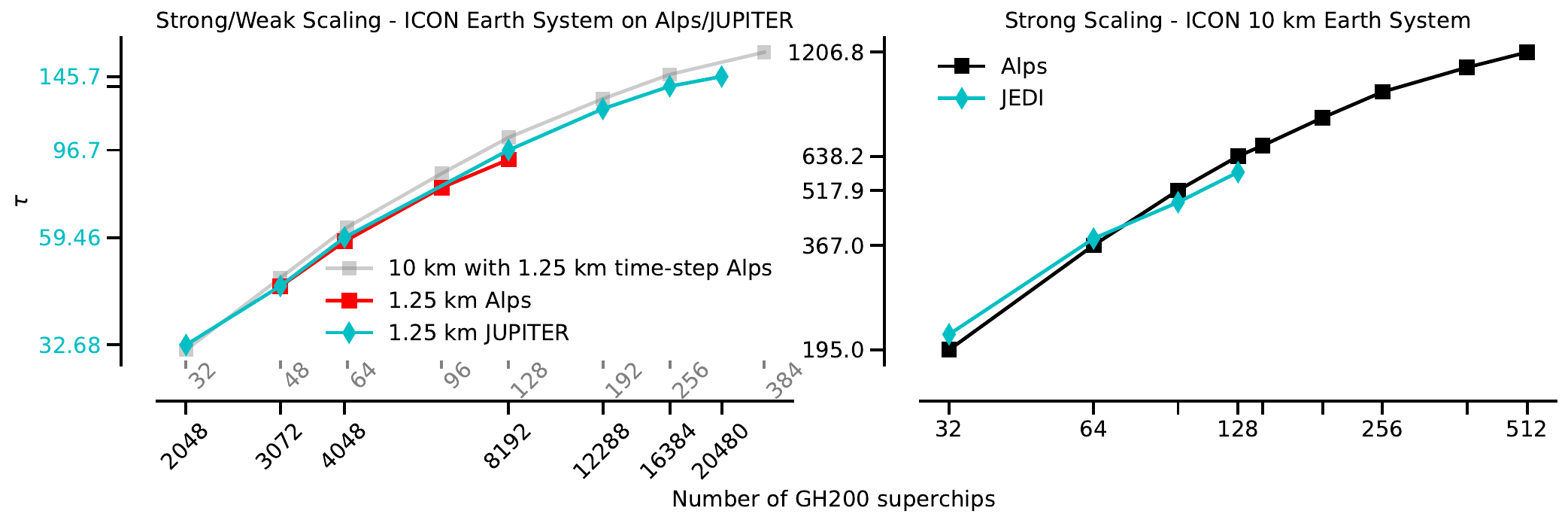}
  \caption{\sisetup{detect-all=true}Strong scaling of the full ICON Earth system model at \qty{1.25}{\kilo\meter} grid-spacing (left). The gray line shows scaling with a configuration with a factor 64 less grid cells (\qty{10}{\kilo\meter} resolution) as the \qty{1.25}{\kilo\meter} configuration but using the same timestep, serving as a reference line assuming perfect weak-scaling. Strong scaling of the ICON Earth system model with \qty{10}{\kilo\meter} grid-spacing on Alps and \jupiter{JEDI} (right)}
  \label{fig:StronAndWeak-Scaling}
\end{figure*}

We performed four sets of benchmark experiments. With our \qty{10}{\kilo\meter} development configuration, we ran a series of simulations for one simulated day each on 32 to 128 GH200 superchips on the \jupiter{JEDI} system to measure strong scaling. The same configuration was also run on Alps, where we extended the strong-scaling measurements to 512 superchips. Results are shown in Figure~\ref{fig:StronAndWeak-Scaling} on the right-hand side. Performance on the two systems is comparable, with slightly better performance on \jupiter{JEDI} for low superchip counts, while the performance on Alps is slightly better for higher superchip counts. The scaling curve flattens when approaching to 512 superchips. At this point only about \num{10800} horizontal grid cells are distributed to each GPU for calculations, which is too little to fully utilize the GPU compute power in our configuration with 90 levels.

To measure weak-scaling, we performed another set of experiments with the \qty{10}{\kilo\meter} configuration on Alps, where we use the same time steps of \qty{10}{\second} as needed for the \qty{1.25}{\kilo\meter} computations. Increasing the grid-spacing of the computational mesh from 10~km to \qty{1.25}{\kilo\meter} grid-spacing corresponds to a 64 fold increase of grid cells in the horizontal (Table~\ref{tab_resolution}). This means, the \qty{10}{\kilo\meter} configuration has the same computational load on 32 superchips, as the \qty{1.25}{\kilo\meter} setup on \num{2048} superchips (Figure~\ref{fig:StronAndWeak-Scaling}, left).

Finally, we also performed scaling simulations with the \qty{1.25}{\kilo\meter} full Earth system configuration of ICON on Alps and JUPITER. The smallest node count we could fit the configuration with its nearly \num{1e12} degrees of freedom (see Table~\ref{tab_resolution}) is \num{2048} superchips of JUPITER ($\qty{\approx 10}{\percent}$ of the full system) achieving a temporal compression of $\tau=32.7$. The production setup scales well to \num{20480} superchips, the maximum number of nodes to which we had access for the strong scaling experiment, achieving a temporal compression of $\tau=145.7$ on JUPITER and $\tau=91.8$ on \num{8192} superchips of Alps (see Figure \ref{fig:StronAndWeak-Scaling}).

Weak scaling results from the benchmarks with the \qty{10}{\kilo\meter} configuration with the time step of the \qty{1.25}{\kilo\meter} configuration (gray curve showing the \qty{10}{\kilo\meter} ICON Earth system model in the left plot of figure \ref{fig:StronAndWeak-Scaling}), gives a weak scaling efficiency of about \qty{90}{\percent} while increasing the size of the computational problem by a factor of 64. For 384 superchips with the \qty{10}{\kilo\meter} resolution (with the \qty{1.25}{\kilo\meter} timestep), we get $\tau\approx167$. Applying this to the \qty{1.25}{\kilo\meter} configuration,  accounting for the \qty{90}{\percent} weak scaling estimated above, gives a $\tau = 150$ for the \qty{1.25}{\kilo\meter} model on \num{24576} superchips (which corresponds to the full JUPITER system).
                    
Our complex Earth system model configuration requires all components to scale similarly well to achieve results up to highest superchip counts of the largest machines.   We have configured the model so that the most computationally expensive component dictates the maximum time compression achievable. At superchip counts not yet available to us on Jupiter, global communication in the 2d solver in the ocean model can become a bottleneck. Performance tests at lower node counts give us confidence that a recent refactoring of this global solver will allow us to maintain scaling of the \qty{1.25}{\kilo\meter} configuration across the entirety of JUPITER.

Since our \qty{1.25}{\kilo\meter} setup runs for 3 simulated hours, we did not write any output during the time loop but wrote restart files consisting of the full model states at the end of the simulation. We measured I/O performance based on the reading and writing of these restart files. For the \qty{1.25}{\kilo\meter} setup the restart files have a size of \qty{9265.50}{\gibi\byte} for the atmosphere and \qty{7030.91}{\gibi\byte} for the ocean respectively.
To read and write the restart files for the ocean part, we ran up to \num{2579} processes when running on \num{8000} superchips. Staggered reading  allows us to read the restart at a rate of \qty{615.61}{\gibi\byte\per\second}. The writing of the ocean restart file at the end of the simulation achieved a rate of \qty{198.19}{\gibi\byte\per\second}.

\section{Implications}

\begin{figure*}[!h]
  \centering
  \includegraphics[ width=0.95\linewidth]{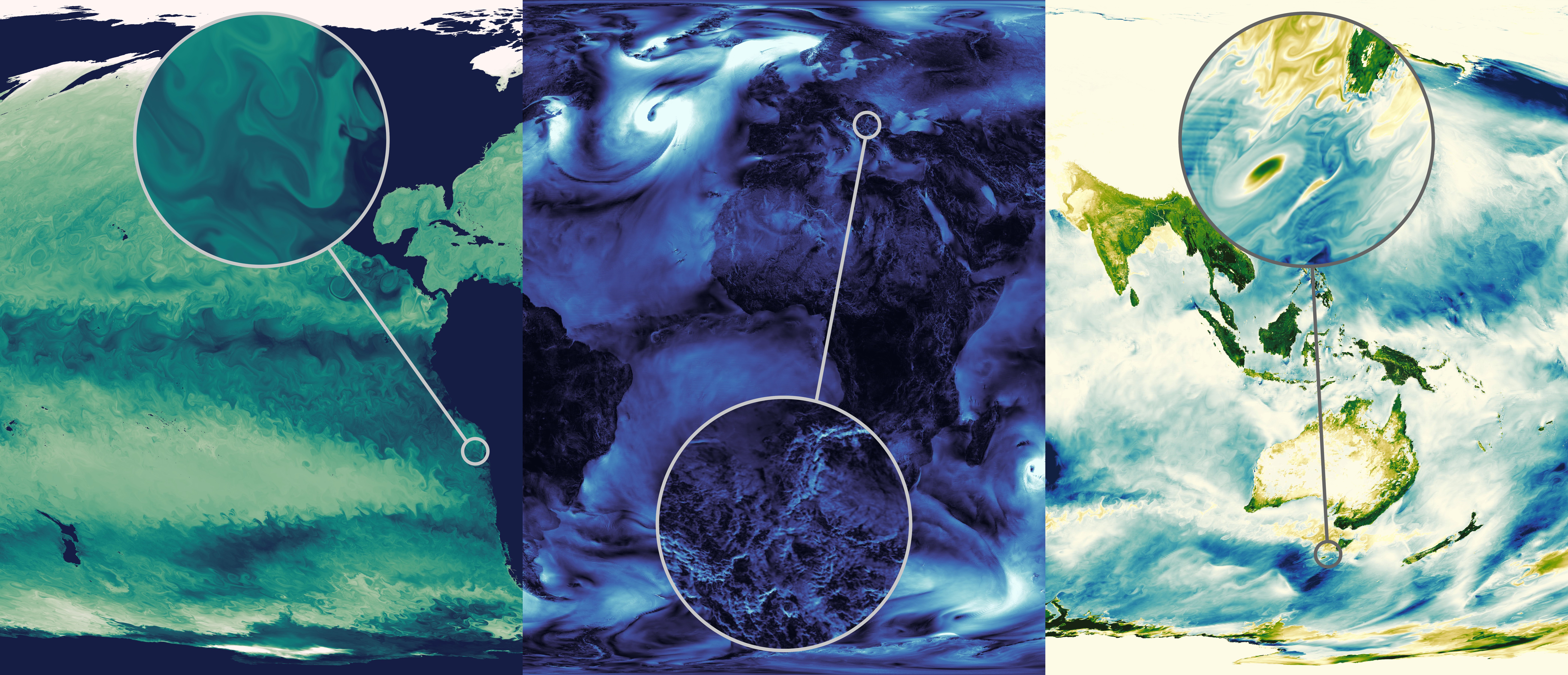}
  \sisetup{detect-all=true}
  \caption{Snapshot of phytoplankton, near-surface wind and air-sea CO$_{2}$ flux at 2020-01-01 03:00. On the left, phytoplankton concentration on a logarithmic scale between \qty{e-9}{\kilo\mol\pascal\per\meter\cubed} and \qty{e-6}{\kilo\mol\pascal\per\meter\cubed} are shown. In the center, surface winds are displayed with color limits in the range from \qtyrange{0}{20}{\meter\per\second}. On the right, the air-sea/land carbon flux is illustrated with the color scale spanning $\pm4\cdot10^{-7}$kg m$^{-2}$\,s$^{-1}$ and green values implying carbon uptake and blue values indicating carbon release for ocean and land (values over the ocean were multiplied by 30 to enhance visibility). 
  Insets highlight specific regions: water mass upwelling off the coast of Chile (left), near-surface wind patterns over the mountains of the Balkans (center), and CO$_2$ flux off the coast of Tasmania (right).}

  \label{fig:pythoplankton}
\end{figure*}

Until now it has not been possible to represent the interaction between the fine and fast scales that mediate the interactions between water and energy -- for instance the daytime heating over land that causes afternoon thunderstorms -- with the large and slow scales of the carbon reservoirs they influence.  These reservoirs collectively define the land biosphere, which then go on to influence how the fast and fine scales develop in the future. Figure~\ref{fig:pythoplankton} shows those Earth system component interactions through the flow of carbon between them.

For the first time, we simulate the impact of small scales on the carbon flows, globally.  More tangibly, this means we represent the global fluxes of carbon as the net result of local interactions between vegetation, topographic features, watersheds, local wind systems, ocean eddies and a realistic representation of the spatial and intensity distribution of precipitation. Capturing these types of interactions require a model with local (km-scale) granularity, globally, an explicit representation of a diverse array of Earth system processes, and a throughput (or time-compression) of about 100. The computational challenge was thought to be unachievable, forcing researchers to focus on approaches that would otherwise be suboptimal (e.g. \cite{Eyring2024}).  We show that this thinking was wrong, and km-scale global models of the full Earth system are realizable on today's technology.

Our computations bring Earth system modeling to the domain of high-performance computing.  Previous and pioneering studies demonstrated the relevance of exascale computing for studies of the physical climate system \cite{SatohEtAl2019,Taylor2023}. This existing capability is helping to uncover new physics, for instance in how imposed changes to the land-surface influence the atmosphere \cite{yoonMutedAmazonRainfall2025}.  By expanding this capability we demonstrate that it can now also be used to study the Earth system in all its fullness.  This opens up new and exciting scientific frontiers.  

One implication of our work is to show that the complexity of the problem, one of the aspects thought to make it difficult, makes it tractable.  This is because it allows us to pair the physical complexity with the complexity of today's computational environments to expose different types of parallelism and use machines more efficiently than for simpler problems. In our case this means using the powerful CPU on the GH200 superchip to couple the rest of the Earth system to the atmosphere, more or less for free.  We also show how a flexible structuring of the solvers, to use GPUs or CPUs or, as in the case of the ocean biogeochemistry, to run concurrently or as part of the main kernels, makes it possible to change the configuration to efficiently adapt to very different architectural profiles.

Our achievement also demonstrates the potential of new approaches to help separate the concerns of domain scientists and computer scientists, which makes it possible for them to work more closely together.  Using the Data Centric programming paradigm we optimize a large kernel of a native Fortran code to run more efficiently on a demanding computational architecture than the originally heavily optimized Fortran implementation instrumented with various libraries.  Our approach preserves the readability of the code, making it easier to maintain and modify and retaining its familiarity for the domain scientist.  Hence, a further implication of our work, beyond showing the relevance of high performance computing to Earth system modeling, is to show the relevance of computer scientists to Earth system scientists.

Looking forward, with the capability already achieved, new machines with architectures similar to those used to perform our benchmarks are coming on line. JUPITER is roughly 2.5 times larger than Alps when at full capacity.  With the performance demonstrated for \num{1024} nodes of JUPITER ($\tau=59.5$, Figure~\ref{fig:StronAndWeak-Scaling}), this amounts to an ability to simultaneously simulate two scenarios of future warming, each for thirty years, with three ensemble members to sample variability, in a little less than half a year. This has enormous and enduring potential to provide \emph{full global Earth system information on local scales} about the implications of future warming for both people and eco-systems, information that otherwise would not exist.

\section*{Acknowledgements}

The authors acknowledge the ICON partnership — including the German Weather Service (DWD), the Max Planck Institute for Meteorology, the German Climate Computing Center (DKRZ), the Karlsruhe Institute of Technology (KIT), MeteoSwiss, and ETH Zurich — for their continued development and improvement of the ICON model. The work of many members of this partnership over the years has laid the foundation for the results presented in this study. The further development of ICON for the efficient use of exascale HPC systems was provided by the German  Federal Ministry of Research, Technology and Space (BMFTR) funded project WarmWorld, under grant numbers 01LK2202, 01LK2203 and 01LK2204 and the BMFTR funded project IFCES2-Scalexa under grant 16ME0692. The Swiss National Supercomputing Centre (CSCS) provided access to the Alps system for scaling runs. This project received access to the JUPITER and the JEDI system through the JUPITER Research and Early Access Program (JUREAP). 

We would like to express our gratitude to Thomas Lippert from JSC for contributing to shaping the project in its very early stages and Thomas Schulthess and Colin McMurtrie of CSCS for providing critical help right before the finish line. Christelle Piechurski and Peter Messmer from NVIDIA supported the project and assisted in getting access for scaling runs. We also like to thank Pay Giesselmann and Julius Plehn from DKRZ for providing energy measurement infrastructure on Levante and helping with the measurements, Tatiana Ilyina of the University of Hamburg for advice in the usage of the ocean biogeochemistry model HAMOCC and Niklas Röber of NVIDIA for visualization suggestions.

The DaCe development for ICON was contributed to by Afif Boudaoud, Berke Ates, Philipp Schaad, Alexandros Ziogas (all at ETH) and was funded by the European Research Council (Project PSAP, No. 101002047), the Swiss State Secretariat for Education, Research and Innovation (SERI) under the SwissTwins project and was partially supported by the ETH Future Computing Laboratory (EFCL), financed by a donation from Huawei Technologies.

\bibliographystyle{plainnat}
\bibliography{ref}

\end{document}